\documentclass{aa}

\usepackage{natbib} 
\usepackage{psfig} 

\begin{document} 
 
\title{Age-Distance diagram for close-by young neutron stars} 
 
\titlerunning{Age-distance diagram} 
 
\author{S.B. Popov 
            \inst{1,2} 
         } 
 
   \offprints{S. Popov} 
 
\institute{Universit\`a di Padova, Dipartimento di Fisica,  
via Marzolo 8, 35131, Padova, Italy 
            \and 
   Sternberg Astronomical Institute, 
Universitetski pr. 13, 119992 Moscow, Russia\\ 
\email{polar@sai.msu.ru} 
}

   \date{} 
 
\abstract{ An age--distance diagram for 
close-by young isolated neutron stars of different types is introduced
and discussed.
It is shown that such a diagram can be a useful tool for studying 
of populations of sources if their 
detectability strongly depends on age and distance.
\keywords{stars: neutron  -- stars: evolution -- pulsars: general}} 
 
\maketitle 

\pagebreak
 
\section{Introduction} 

 There is a growing evidence that young neutron stars (NSs) can have 
a wide diversity of parameters and correspondently can manifest themselves
as sources with different properties: normal radio pulsars (PSRs), 
soft-gamma repeaters (SGRs), anomalous X-ray pulsars (AXPs), radioquiet
X-ray bright central compact objects in supernova remnants (CCOs in SNRs),
$\gamma$-ray (for example EGRET) sources,
radioquiet dim X-ray sources (like the Magnificent Seven, hereafter M7).
In addition to all these stocks of sources there can be young NSs which are
so dim in any band to avoid detection with present day instrumentation.

 If a NS is young enough ($\la 10^5$--$10^6$~years) then as a ``bonus''
we have a possibility to detect its thermal emission.
Such observations are of great importance 
because studying of the thermal evolution
is one of a very few chances ``to look inside'' a NS as far as the thermal
evolution strongly depends on the internal structure.
Also sometimes thermal emission is the only way to detect a young NS.

 Fortunately the solar vicinity ($\la 0.5$--1~kpc) is enriched with
young ($\la$ few Myr) NSs due to  the Gould Belt
which makes studying of young NSs easier. 
In particular NSs from the Gould Belt 
can be responsible for a significant part of unidentified EGRET sources
(see Grenier 2000\nocite{g2000}, Gehrels et al. 2000\nocite{geh2000}) 
and for the M7 (Popov et al. 2003\nocite{ppc03}).

Young close-by NSs are now actively studied both from observational and 
theoretical sides (see recent reviews in Kaspi et al. 2004\nocite{krh04},
Popov \& Turolla 2004\nocite{pt04} and references there in). 
In this short note I introduce and discuss a simple
age-distance diagram, which can
help to understand properties of the local population of young NSs.

\section{Age-distance diagrams} 

 If we are speaking about observations of thermal emission of young 
cooling NSs then most of their properties depend on their age, and
their detectability in X-rays of course strongly depends on distance 
(not only because of flux dilution, but also because of strong 
interstellar absorption of soft X-rays).
In that sense it is useful to plot an age-distance diagram (ADD) for these
objects.

There are several reasons for introduction of ADDs.
At first an ADD easily illustrates distributions of sources in age and 
distance. Then it is possible to plot ``expectation lines'' 
for abundance of sources of different types and compare them with data.
Finally since observability of sources depends mainly on their ages
and distances, it is possible to illustrate observational limits.

In the following figures ADDs for close-by young NSs are presented.
These objects can be observed in soft X-rays due to their thermal emission.
The thermal evolution strongly depends on mass. According to different models
(see Kaminker et al. 2002\nocite{kyg02}, Blaschke et al. 2004\nocite{bgv04}
and references there in) a NS with a mass $\sim 1.3$--$1.4\, M_{\sun}$
remains hot ($T\sim (0.5$--$1)\, 10^6$~K, or equivalently $\sim$ 50--100 eV) 
up to $\sim$~0.5--1 Myr. 
It corresponds to a luminosity $\sim10^{32}$~erg~s$^{-1}$.
Usually NSs with low masses (1--1.3 $M_{\sun}$) cool down slower,
with higher masses~--~faster. 

To plot an ADD it is necessary to fix maximal values for age and distance
for selection of sources.
As a limiting age for selection of observed sources
I choose 4.25 Myrs. This is the time after which even a 
low-mass NSs cools down to $\sim 10^5$~K and becomes nearly undetectable
in X-rays
(see Kaminker et al. 2002\nocite{kyg02}, Popov et al. 2003\nocite{ppc03}).
A limiting distance is taken to be equal to 1 kpc (because of absorption
it is difficult to detect in X-rays an isolated NS at larger distance). 
In the table 1 in \cite{ppc03} 
there are 20 sources of different nature which are supposed to be
young close-by NSs (age $<$ 4.25 Myr, distance $<$ 1 kpc): 
the M7, Geminga and the geminga-like source RX J1836.2+5925,
four PSRs with detected thermal emission (Vela, PSR 0656+14, PSR 1055-52,
PSR 1929+10\footnote{It should be mentioned, that soft X-ray emission of 
PSR 1929+10 can be due to polar caps or due to non-thermal mechanism}), 
and seven PSRs without detection of thermal emission.
Not for all of these sources 
there are good estimates of ages and/or distances
(especially for the M7). 
In the figures there are data points for 13 objects 
for which such estimates exist (distance to PSR B1055-52 is uncertain, 
and we accept it to be 1 kpc).
Note, that there are also two PSRs with distance $<$1~kpc and with ages
in between 4.25 and 5 Myr (PSR B0823+26, PSR B0943+10) and PSR J0834-60
with distance 0.49~kpc and unknown age. These three objects
are not included into the figures. Also  PSR B1822-09
with age 0.23~Myr and distance 1~kpc is not plotted.

Two types of objects are distinguished on the graphs: 
detected and undetected due to
thermal X-ray emission (remember about seven additional sources -- 
six from the M7 and one geminga-like object -- 
for which there are no definite determinations of age or/and distance). 

I start with a simple toy-model plot, then a realistic initial distribution
of NSs is considered, and finally dynamical effects are taken into account.

In the first figure a simplified example of an ADD is shown. 
This example is very illustrative as far as 
all dependences in the limits of small an large distances are clear.
Here it is  assumed 
that the Sun is at the center of a spherically symmetric
structure with $R_\mathrm{b}=300$~pc 
(in some sense it mimics the Gould Belt).
Then at larger distances NSs are considered to be born in a disk.
NS formation rate in {\it "The Belt"} is assumed to be 
$\dot n_1=$235~Myr$^{-1}$~kpc$^{-3}$
(which corresponds to 26--27 NSs in one Myr up to $R_\mathrm{b}$). 
In the disk the same quantity
is $\dot n_2=$10~Myr$^{-1}$~kpc$^{-2}$ 
(280 NSs in one Myr up to 3 kpc). This value are more
or less equal to the one used in \cite{ptp04}.
So at small distances ($R<R_\mathrm{b}$) number of sources grows nearly as
the cube of distance, and at large distances -- as the square.

The solid line (the lower one) corresponds to one object of given age
at a specified distance. The line is calculated as:

$$
t=\frac1{4/3 \pi \dot n_1 R^3}, \, R<300\,{\rm pc};
$$ 

$$
t=\frac1{4/3 \pi \dot n_1 R_b^3 + \pi \dot n_2 (R-R_b)^2} , \, R>300\,{\rm
pc}.
$$
Here $t$ is in Myrs, $R$ -- in kpc.

The dotted line  corresponds to
13 objects (there are 13 sources shown as symbols).
Three lines (dashed, dot-dashed and dot-dot-dashed)
corresponds to one sources from one of the three mass
ranges: 1.05--1.3~$M_{\sun}$ (73\% of all NSs), 
1.3--1.55 (26\%), 1.55--1.8 (1\%). They were obtained just by shifting 
the solid line along the vertical (age) axis for factors 
$\sim$1.37, $\sim$3.85, 100. These values correspond to the NS mass spectrum
which was used in \cite{ptp04}.   
All five described lines have the same shape which is better seen in
the top one (dot-dot-dashed).

In addition I add a "visibility" line (the solid one in the middle of the
plot). The idea is to show a maximal distance for a given age (or vice versa
a maximal age for a given distance) at which a hot (i.e. low-mass) NS can be 
detected.
The cooling curve is taken from \cite{kyg02} for 
$M=1$--$1.3\, M_{\sun}$ (in the model of these authors cooling of NSs with
$M\la 1.35\, M_{\sun}$ is nearly mass-independent).
Such curves were used for example in (Popov et al. 2003\nocite{ppc03}). 
As soon as a cooling curve is fixed then the age determines the luminosity
of the object.
The limiting unabsorbed 
flux is assumed to be $10^{-12}$~erg~cm$^{-2}$~s$^{-1}$.
According to 
WebPIMMS\footnote{http://heasarc.gsfc.nasa.gov/docs/corp/tools.html} 
it corresponds to $\sim0.01$ 
ROSAT PSPS counts per second
for $N_H=10^{21}$~cm$^{-2}$ and a blackbody spectrum with $T=90$~eV,
or to $\sim0.1$ 
ROSAT PSPS counts per second
for $N_H=10^{20}$~cm$^{-2}$ and a blackbody spectrum with $T=50$~eV.
The latter values corresponds to the dimmest source among  the
M7 -- RX J0420.0-5022;
the former to possibly detectable hot far away objects.
Without any doubt such simple approach
underestimates absorption at large distances. So
the age at which a NS is still observable is overestimated, but for
distances $\la$~1~kpc and ages $\la$~1~Myr 
it should not be a dramatic effect.

\begin{figure}
\vbox{\psfig{figure=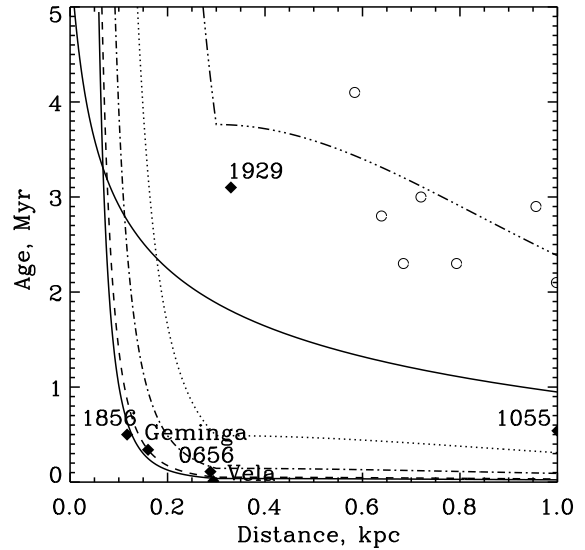,width=\hsize}}  
\caption[]{  Distance-age diagram for a toy-model. 
Filled symbols -- sources detected in X-rays,
empty -- non-detected.
} 
\label{fig:primer2} 
\end{figure} 

\begin{figure}
\vbox{\psfig{figure=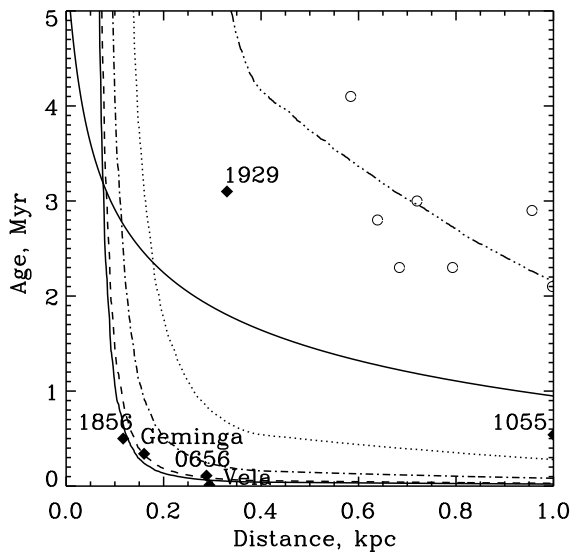,width=\hsize}}  
\caption[]{Age-distance diagram for close-by young NSs with realistic
birth rate and initial distribution, but without 
dynamical effects.
} 
\label{fig:data} 
\end{figure}

In the figure \ref{fig:data}
I present the same plot as in figure
\ref{fig:primer2} but for a  more realistic model.
Here NS formation rate is taken from the numerical model used in 
\cite{ptp04}.
NSs were not followed during their dynamical evolution, so numbers
correspond to initial values (this is equivalent to zero kicks and progenitor
velocities). In 4.3 Myr in 1 kpc around the Sun about 200 NSs are expected
to be born.

In the final picture an ADD with an inclusion of the dynamical evolution
of NSs is shown. I.e. here  NSs' movements in the 
Galactic potential are accurately calculated 
(all procedures are the same as in Popov et al. 2004\nocite{ptp04},
kick velocities are taken from Arzoumanian et al. 2002\nocite{acc2002}).
For clearity in the last figure the names of objects are not plotted.
Five solid 
lines are plotted for 1, 4 (it should be close to the line for 
NSs in the mass range 1.3--1.55 $M_{\sun}$, see above), 
13, 20 and 100 sources for comparison with the 
previous figures. Obviously all lines are shifted to the left
in comparison with fig.\ref{fig:data} because
sources are leaving the volume presented on the graphs ($R<$~1 kpc).
Shifts are more pronounced for large expected numbers of NSs.
The dotted line represents the ``visibility'' line.
 
\begin{figure}
\vbox{\psfig{figure=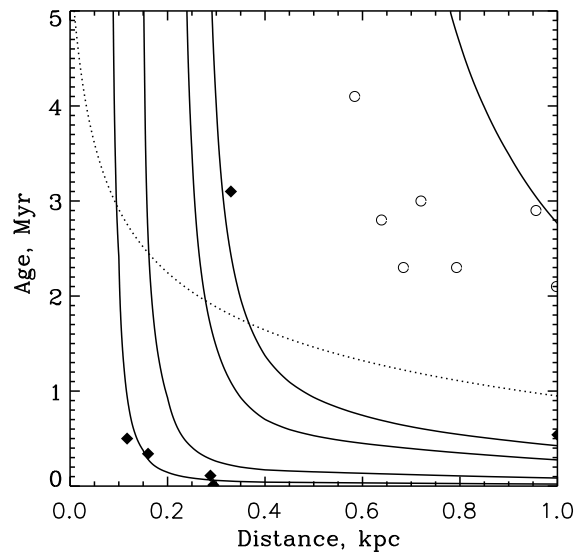,width=\hsize}}  
\caption[]{ Age-distance diagram with dynamical effects. 
Solid lines (from bottom to the top) corresponds to 1, 4, 13, 20 and
100 sources. The dotted line is the ``visibility'' line (see the text).
} 
\label{fig:run} 
\end{figure} 


\section{Discussion and conclusions} 

The fact that the line for 20 sources 
lies below  $\sim$1/2 of the observed
points tells us that according to our model not all NSs are observed.
For example, at R=1~kpc we expect to see 20 sources with ages $\sim$0.5~Myr,
but in reality we observe 
20 sources with age $<$4.25~Myr. It means that for constant NS
formation rate we see only about 10\% of them. 
One has to bear in mind that for
seven  sources ages or/and distances are unknown, 
but they are suspected to be young and close, 
this is why here we discuss 20 and not 13 sources. 
Their inclusion or exclusion 
changes our conclusions quantitatively, but not qualitatively.
Six of the M7 sources which are not plotted on the graphs due to lack of
good distance measurement 
should populate the left bottom corner of the graphs,
since according to models of NS thermal evolution (see for example
Kaminker et al. 2002\nocite{kyg02}, Blaschke et al. 2004\nocite{bgv04})
their ages are expected to be $\la$~1~Myr, and their distances should be
$\la$~500~pc.
NSs can easily escape detection as {\it coolers} simply because
they cooled down in $\sim$~1~Myr. A fraction of PSR can be undetected
due to beaming (however, they may be observed as EGRET sources).

PSR B1929+10 lies above the "visibility line".
However it is unclear if X-ray emission of this faint 
(0.012 ROSAT cts~s$^{-1}$)
source is due to non-thermal mechanism or not (Becker \& Tr\"umper 
1997\nocite{bt97}). \cite{ll1999} 
suggested an additional heating for this PSR due to 
some internal mechanisms. Now this object is not widely accepted as a {\it
cooler} and usually it is not plotted on the $T-t$ graphs with cooling
curves (see also recent papers by Zavlin \& Pavlov 2004\nocite{zp04} 
and Becker et al. (2004)\nocite{bwt04} where
the authors briefly comment on that source).

Surprisingly fluctuations in the part of the plot
with small number of NSs are not large.
Probably we see nearly all low-massive NSs with ages $\la$~1~Myr at
$R\la$~300--400~pc.
It should be noted that we are very lucky to have such a young and close
object as the Vela pulsar. If one believes in the mass spectrum with small
fraction of NSs with $M>1.5\, M_{\sun}$ then inevitably one has to conclude
that Vela cannot be a massive NS. It can be important in selection of
cooling models (see Blaschke et al. 2004\nocite{bgv04}). If in a model
Vela is explained only by a cooling curve for  $M>1.5\, M_{\sun}$ then
the model may be questionable. In fact, having such a young and
 massive NS so close is very improbable.
Similar conclusion can be made for Geminga and PSR 0656+14.

There is a deficit of sources below the "visibility" line.
These sources could be already detected as dim X-ray sources by ROSAT 
but were not identified as isolated NSs.
The limit for the number of isolated NSs
from the BSC (Bright Source Catalogue) is
about 100 sources at ROSAT count rate $>$0.05 cts~s$^{-1}$ 
(see Rutledge et al. 2003\nocite{rut2003}). However we do not expect
to see that many young isolated NSs due to their thermal emission.
An expected number of  {\it coolers} observable by ROSAT
is about 40 objects with ages $\la $~1 Myr inside
$R\la 0.5$--1 kpc (see the ``visibility'' line in the
fig.\ref{fig:run} in comparison with the line for 20 sources, for example).
As it was shown by \cite{ptp04} most of unidentified {\it coolers} in ROSAT 
data are expected to be located at low galactic latitudes in crowded regions.

In the figure \ref{fig:data} and especially in the figure \ref{fig:run}
in comparison with the first graph it is visible that lines rise
at small distance faster than the third power of distance.
There are two reasons for such behaviour.
The first one is related to the initial spatial distribution of NSs
(or in another way to the accepted distribution of progenitors, see
details in Popov et al 2004\nocite{ptp04}): no NSs are supposed to born
in $\sim 50$--100~pc around the Sun (depending on direction).
The second reason (which is the main one)
is connected with high spatial velocities of NSs.
With $V\sim 300$~km~s~$^{-1}$ a NS makes 300 pc in $10^6$ years.
As a result the number of NSs with ages $\ga$~1 Myr in the solar proximity
is decreased.

According to the classical picture all NSs were assumed to be born
as PSRs more or less similar to Crab. Depending on the initial parameters
they were assumed to be active for $\sim 10^7$~yrs.
In the last  years this picture was significantly changed 
\cite{gv00}. Without any doubts we see a deficit of young
PSRs in the solar vicinity
in comparison with an expected value of young NSs from the Gould Belt and
the beaming factor cannot be the only reason for this deficit.  
This deficit  can be explained if one assumes
that a significant part of NSs do not pass through the radio  pulsar stage
or that this stage is extremely short for them.
Of course fluctuations (in time and space) of NS production rate 
can be important as far as statistics is not that high.

Comparison of fig.\ref{fig:data} and \ref{fig:run} shows that influence
of dynamics is important mostly for old or far away stars.
Since the kick velocity distribution is bimodal with 40\%
of ``low-velocity'' ($\la$~300~km~s$^{-1}$) and 60\%
of high-velocity ($\ga$~300~km~s$^{-1}$) NSs 
(Arzoumanian et al. 2002\nocite{acc2002}),
the role of the dynamical evolution is more important for objects from the 
high-velocity peak of the distribution. Obviously in that case we expect to 
see more middle-age (1--3 Myr) 
objects from the low-velocity peak closer to us: among 20 objects
only PSR B2045-16 with
transverse velocity $V_\mathrm{t}\sim350$~km~s~$^{-1}$ belongs to the
high-velocity peak (for several objects velocity is unknown).
It is unclear if such effect can lead to any observational consequencies.
It is quite possible that compact stars populating the two parts
of the velocity distribution can have different properties. 
Even in the 
absence of correlations of some parameters with velocity {\it inside} each
part of the kick distribution it can exist between the two parts
(see Bombaci \& Popov\nocite{bp04} for a more detailed discussion).
For example as far as kick is definitely connected with the physics of a SN
explosion, then if a significant part of mass is obtained by a NS due to 
fall-back, we can expect if not a mass-velocity correlation then at least
a difference in an average mass in the low- and high-velocity peaks 
(see Popov et al. 2002\nocite{popov2002} on correlations with mass).
As far as the cooling history is determined by a NS mass, then a correlation
between temperature (for a given age) and velocity can be expected.

I conclude that an ADD can be a useful tool for illustration of the
properties of close-by NSs. Its modifications can be applied to other types
of sources. For example an addition of the third axis 
(for $p$ or $\dot p$ for example) 
can be useful in discussing the population of radiopulsars. 

\begin{acknowledgements} 
It is a pleasure 
to thank Ignazio Bombaci and Hovik Grigorian 
for numerous discussions. Special thanks to 
Roberto Turolla for carefull reading of the text and comments,
and to Valentina Bianchin for 
assistance with IDL plots and for permanent support.
\end{acknowledgements} 
 

\end{document}